\documentclass[%
 twocolumn,
superscriptaddress,
showpacs,preprintnumbers,
nofootinbib,
 amsmath,amssymb,aps,
prd,
floatfix,
]{revtex4-1}

\usepackage{lipsum} 

\usepackage{bm} 
\usepackage{physics}
\usepackage{comment}
\pdfoutput=1
\usepackage{amsmath,amsfonts,amsthm}
\usepackage{esdiff} 
\usepackage{booktabs}  
\usepackage{url}  
\usepackage[hidelinks]{hyperref}  
\usepackage{relsize}

\usepackage[caption=false]{subfig}

\usepackage[normalem]{ulem}

\usepackage[usenames,dvipsnames,svgnames,table]{xcolor}

\usepackage[export]{adjustbox}
\usepackage{graphicx}

\usepackage[sc]{mathpazo} 
\usepackage[T1]{fontenc} 
\usepackage{microtype} 

\usepackage{booktabs} 
\usepackage{float} 
\usepackage{hyperref} 

\usepackage{lettrine} 
\usepackage{paralist} 

\usepackage{makecell} 

\usepackage{titlesec} 
\renewcommand\thesection{\Roman{section}} 
\renewcommand\thesubsection{\Alph{subsection}} 
\titleformat{\section}[block]{\large\scshape\centering\bfseries}{\thesection.}{1em}{} 

\titleformat{\subsection}[block]{\scshape\centering}{\thesubsection.}{1em}{} 

\DeclareCaptionFormat{myformat}{#1#2#3\hrulefill}
\begin{document}

\title{Constraining solar electron number density via neutrino flavor data at Borexino}

\author{Caroline Laber-Smith}\thanks{Corresponding author}
\email{labersmith@wisc.edu}
\affiliation{Department of Physics, University of Wisconsin: Madison, Madison, WI 53706, USA}
\author{Eve Armstrong}
\email{earmst01@nyit.edu}
\affiliation{Department of Physics, New York Institute of Technology, New York, NY 10023, USA}
\affiliation{Department of Astrophysics, American Museum of Natural History, New York, NY 10024, USA}
\author{A. Baha Balantekin}
\email{baha@physics.wisc.edu}
\affiliation{Department of Physics, University of Wisconsin: Madison, Madison, WI 53706, USA}
\author{Elizabeth K. Jones}
\affiliation{Department of Physics, Harvey Mudd College, Claremont, CA 91711}
\author{Lily Newkirk}
\affiliation{Department of Physics, New York Institute of Technology, New York, NY 10023, USA}
\author{Amol V.\ Patwardhan}
\email{apatward@umn.edu}
\affiliation{School of Physics \& Astronomy, University of Minnesota, Minneapolis, MN 55455}
\affiliation{SLAC National Accelerator Laboratory, 2575 Sand Hill Rd, Menlo Park, CA 94025, USA}
\author{Sarah Ranginwala}
\affiliation{Department of Physics, New York Institute of Technology, New York, NY 10023, USA}
\author{M. Margarette Sanchez}
\email{msanch11@nyit.edu}
\affiliation{Department of Physics, New York Institute of Technology, New York, NY 10023, USA}
\author{Hansen Torres}
\affiliation{Department of Physics, New York Institute of Technology, New York, NY 10023, USA}

\date{\today}

\begin{abstract}
Understanding the physics of the deep solar interior, and the more exotic environs of core-collapse supernovae (CCSN) and binary neutron-star (NS) mergers, is of keen interest in many avenues of research.  To date, this physics is based largely on simulations via forward integration.  While these simulations provide valuable constraints, it could be insightful to adopt the "inverse approach" as a point of comparison.  Within this paradigm, parameters of the solar interior are not output based on an assumed model, but rather are inferred based on real data.  We take the specific case of solar electron number density, which historically is taken as output from the standard solar model.  We show how one may arrive at an independent constraint on that density profile, based on available neutrino flavor data from the Earth-based Borexino experiment.  The inference technique's ability to offer a unique lens on physics can be extended to other datasets, and to analogous questions for CCSN and NS mergers, albeit with simulated data.

\end{abstract}

\maketitle

\section{INTRODUCTION}
{The dynamics of the solar interior is of keen interest in a variety of research areas, particularly the fundamental physics of stellar evolution \cite{Gann:2021ndb}.  To date, constraints on solar properties have come largely from the standard solar model \cite{Vinyoles:2016djt}.  In this scenario, parameters governing solar dynamics are estimated via forward integration based on a reasonable assumed initial equation-of-state \cite{Morel:1999jz}.  While these constraints are the best we have to date, it would be useful to identify an independent lens through which to obtain estimates for comparison.

Inference may provide such a lens.  Inference, also called the "inverse approach" and related to machine learning methodology, is a means to obtain estimates on unknown model parameters based not on \textit{a priori} physical assumptions, but rather on real available data where the data are assumed to arise from the model dynamics.  It is a fundamentally different framework compared to forward integration, and is formulated mathematically to address problems that forward integration is ill-equipped to handle~\cite{armstrong2017optimization}.

We take as our parameter of interest the radially-varying solar electron number density $n_e(r)$.  We are interested in this quantity as it has important implications for neutrino flavor physics.  Standard simulations of solar neutrino spectra are initial-value problems (IVPs) that take $n_e(r)$ as input, where $n_e(r)$ has been calculated by the standard solar model, given observed surface abundances.  In a previous paper~\cite{laber2023inference} we assayed whether that $n_e(r)$ profile was consistent with available Earth-based neutrino flavor data from the Borexino~\cite{BOREXINO:2018ohr} and SNO~\cite{SNO:2009uok} experiments.  We found that it was but that result begged the followup question: is that particular solution for $n_e(r)$ necessarily the only profile that is consistent with the available data?  That is: might the data further constrain, or broaden, the range of permitted density profiles?  Thus was born our choice to reverse the scientific question: can the neutrino data yield an independent constraint on  $n_e(r)$?  

That is the quest in this current manuscript.  Specifically, we seek to estimate $n_e(r)$ directly from neutrino data using not an IVP but rather a two-point boundary-value-problem (BVP) formulation\footnote{We should point out that although applying the SDA method to this problem is a new approach, the idea of estimating solar density through neutrino observations itself has been considered before using other approaches \cite{Balantekin:1997fr,Lopes:2013nfa}.}.  
The specific inference procedure we adopt is statistical data assimilation (SDA).  SDA is an inference methodology invented to extract information from measurements for the purpose of numerical weather prediction~\cite{kimura2002numerical,kalnay2003atmospheric,evensen2009data,betts2010practical,whartenby2013number,an2017estimating}.  It has since gained traction in neurobiology~\cite{schiff2009kalman, toth2011dynamical,kostuk2012dynamical,hamilton2013real,meliza2014estimating,nogaret2016automatic,armstrong2020statistical}.  The aim of SDA is to incorporate information contained in measurements directly into a model, to estimate unknown parameters and the dynamics of the model state variables, both measured and unmeasured.  SDA was invented specifically for the case of sparse data, which is a key feature of our problem; we can measure only a fraction of the neutrinos emitted from the Sun, and at just one location i.e., Earth.  We have used SDA in previous papers~\cite{armstrong2022inferenceA,armstrong2022inferenceB,armstrong2017optimization,armstrong2020inference,rrapaj2021inference} on collective neutrino flavor oscillations in core-collapse supernovae (CCSN), where the aim was to examine nonlinearities of the model through the inverse-problem lens\footnote{To our knowledge, the only other representation of SDA within astrophysics is limited to one group studying the solar cycle~\cite{Kitiashvili2008,kitiashvili2020}, although inference for pattern recognition has grown popular for mining large astronomical data sets~\cite{sen2022astronomical}.}.

Our specific task for SDA is to take Earth-based electron-flavor survival probabilities~\cite{BOREXINO:2018ohr} together with a model of flavor evolution assumed to underlie those data, to infer (i) the electron density $n_e$ at the solar center, and (ii) whether the data show preference for a particular analytical form for the full trajectory $n_{e}(r)$.}

\section{INPUT} 

\subsection{\textbf{Model of neutrino flavor evolution}} \label{sec:evol}

To model the flavor evolution of neutrinos inside the Sun, we follow the same framework as \cite{laber2023inference}. We consider mixing between $\nu_e$ and $\nu_x$ (a superposition of $\nu_\mu$ and $\nu_\tau$) in a two-flavor framework with a mixing angle $\theta \simeq \theta_{12}$ and $\theta_{13} \simeq 0$. All neutrinos are assumed to be produced in the center of the Sun in purely $\nu_e$ flavor states, and subsequently, they experience flavor evolution through a combination of vacuum oscillations and matter effects inside the Sun. Sufficiently energetic neutrinos undergo a mass-level crossing, i.e., the Mikheyev-Smirnov-Wolfenstein (MSW) resonance, and consequently experience enhanced coherent $\nu_e$-$\nu_x$ transformations as they pass through the solar envelope.

The neutrinos may be represented using a polarization vector (i.e., a Bloch vector) with real-valued components, which are related to the wave function amplitudes $\psi_e = \braket{\nu_e}{\psi}$ and $\psi_x = \braket{\nu_x}{\psi}$  in the following manner:
\begin{equation}
\label{polvec}
\vec{P} = \left(
    \begin{array}{c}
      P_x   \\
      P_y   \\
      P_z
    \end{array} \right) = \left(
    \begin{array}{c}
    \psi_e \psi_x^* + \psi_e^* \psi_x \\
    i( \psi_e \psi_x^* - \psi_e^* \psi_x) \\
         |\psi_e|^2 - |\psi_x|^2 \\
             \end{array} \right) .
\end{equation}

In particular, the $\hat{z}$ component $P_z$ of the neutrino polarization vector denotes the net flavor content of that particular neutrino ($e$-flavor minus $x$-flavor), and is related to the electron-flavor survival probability $P_{ee}$ as $P_z = 2P_{ee} - 1$. Since the neutrinos are assumed to be produced as pure $\nu_e$ states in the center of the Sun, we have the initial condition $P_z = 1$ and $P_x = P_y = 0$ for each neutrino. The dynamical equation for neutrino flavor evolution, when expressed in this language, assumes the form of a spin-precession equation:
\begin{equation} 
\label{eq:model1}
\diff{\vec{P}}{r} = \left(\omega \vec{B} + V(r) \hat{z}\right) \times \vec{P},
\end{equation}
The external field driving this precession in flavor space consists of a \lq\lq vacuum\rq\rq\ part $\omega \vec{B}$, and a \lq\lq matter\rq\rq\ part $\vec{V} = V(r) \hat{z}$. $\omega = \delta m^2/(2E)$ is the vacuum oscillation frequency of a neutrino with energy $E$ and mass-squared difference $\delta m^2$.  The unit vector $\vec{B}=\sin(2 \theta) \hat{x} -\cos(2 \theta) \hat{z}$ points along the direction of the $\ket{\nu_2}$ mass eigenstate in flavor space. The function $V(r)$ is proportional to the electron number density, as described in Sec.~\ref{sec:ne}.  
A neutrino propagating out through the Sun encounters a mass-level crossing (MSW resonance) when $V(r) - \omega \cos{2\theta} = 0$, i.e., when the $\hat z$ component of the external field vanishes.

We take our domain of state-variable evolution to span from the center of the Sun to a radius of \({R_\odot}/{2}\) (half the solar radius), where the matter density is sufficiently low to for it to be considered the vacuum regime (i.e., $V(r) \ll \omega \cos{2\theta}$). To connect the state variable evolution within the Sun to measurements of neutrino flavor at the earth, however, requires some careful consideration. One has to take into account that the neutrinos kinematically decohere as they propagate over sufficiently long distances, i.e., the $\nu_1$ and $\nu_2$ mass eigenstates become separated in space as they are moving at slightly different speeds. As a result, the detector \lq\lq catches\rq\rq\ only one mass eigenstate at a time, and not a coherent superposition of both. Taking this effect into account leads to the following transformation between the polarization vector components at $r_\text{final} = R_\odot/2$ -- the outer endpoint of our domain,
and the neutrino electron-flavor survival probability measured on Earth, $P_{ee,\oplus}$~\cite{laber2023inference,Dighe:1999id}:
\begin{equation}
\label{eq:connection}
\left(1-\sin^2 2\theta\right) P_{z,\text{final}} - \cos2\theta \sin2\theta P_{x,\text{final}} = 2 P_{ee,\oplus}-1.
\end{equation}

{ A derivation of the above relation can be found in Appendix~\ref{app:decoh}. The same relation can, in principle, be derived by taking the flavor composition (a $\{P_x,P_y,P_z\}$ triad) at the surface of the Sun, translating it to earth, and averaging over either the neutrino energy or propagation distance (averaging over the product $\omega L$). To derive the same expression using averaging arguments, rather than decoherence, we refer the reader to our previous paper~\cite{laber2023inference}, equations (5)--(12).}

\subsection{\textbf{Model for electron number density $n_e$}} \label{sec:ne}

The matter potential $V(r)$ of Eq.~\ref{eq:model1} is related to the electron number density \(n_e\) as:
\begin{equation*}
V(r)=\sqrt{2}G_F n_e\left(r\right).
\end{equation*}
This term embodies the neutrino mass Wolfenstein correction~\cite{wolfenstein1978neutrino} since it arises from neutrino coherent forward scattering on the background electrons. 

We chose two distinct forms for the matter potential, to check for invariance of results across the two.  Our motivation was as follows.  If the change in \(V(r)\) is slow enough to allow for adiabatic evolution, the resulting neutrino state would depend solely on the matter potential at the endpoints, \(V_0\) and \(V_f\), and not strongly on the shape of \(V(r)\) itself.  In choosing two versions for \(V(r)\), we aimed to test this expectation.

The first version for \(V(r)\) was an exponential decay:
\begin{equation}\label{eq:expdecaymodel}
V(r) = V_0 \exp{-\frac{2r}{R_\odot}\ln\frac{V_0}{V_f}}.
\end{equation}
This form can approximate the decay of the matter profile of the standard solar model near the edge of the Sun, while remaining analytically simple.

The second version was a logistic function:
\begin{equation}\label{eq:logmodel}
    V(r) = V_0 \left[\frac{\left(e^{k\frac{R_\odot}{2}}-1\right)\left(\frac{V_f}{V_0}\right)^q}{\left(e^{k\frac{R_\odot}{2}}-e^{kr}\right)\left(\frac{V_f}{V_0}\right)^q - 1 + e^{kr}}\right]^{1/q}
\end{equation}
\noindent Here, \(q = 1.7\) and \(k = 1.35\times 10^{-5} \text{ km}^{-1}\). We adopted this form so that, together with the chosen parameter values, it would more closely match the shape of the matter potential from the standard solar model, compared to the simpler exponential form.  Namely, the falloff near the solar center is shallower.

Both models were designed so that altering the value of \(V_0\) would keep \(V\left({R_\odot}/{2}\right) = V_f\) fixed, at a value near zero, in agreement with the standard solar model~\cite{bahcall2005new}.  The aim of this paper, to be described in Sec.~\ref{sec:expers}, was to estimate from data the value of $V_0$ -- the matter potential at the solar center.  (Model quantities taken to be known are listed in Table~\ref{table:model}.)
\setlength{\tabcolsep}{5pt}
\begin{table}[htb]
\caption{Model quantities taken to be known.  The $V_0$ values here refer to the initial (proof-of-concept) stage of testing the procedure, prior to parameter estimation (see Sec.~\ref{sec:expers}).}
\vskip 0.2cm
\centering
\begin{tabular}{| c | c |} \toprule
\hline
 \textit{Parameter} & \textit{Value [unit]} \\\midrule \hline
 $R_{\odot}$ & \(6.957\times 10^{5} \text{ km}\)\\
 $\theta$ & 0.5838 rad \\ 
 $G_F$ & \(1.166 \times 10^{-11} \text{ MeV}^{-2}\) \\
 $\delta m^2$ & \(7.53\times10^{-17} \text{ MeV}^{2}\) \\\hline
 $V_0$ values & \makecell{\(0.015,0.02,0.025,0.03,0.035,\)\\\(0.055,0.075,0.095 \text{ km}^{-1}\)}\\
 \bottomrule \hline
\end{tabular} 
\label{table:model}
\end{table}

\subsection{\textbf{Data}}

{ We used $^8$B daytime neutrino flux observed by the Borexino \cite{BOREXINO:2018ohr} experiment, which provides electron flavor survival probabilities across different energy bins\footnote{ 
We chose not to include other neutrino experiments {(SNO~\cite{SNO:2011hxd} or Super-Kamiokande~\cite{Super-Kamiokande:2016yck}), since they present the neutrino data in terms of a fitting function for electron flavor survival probability vs energy, rather than as survival probabilities in specific  energy bins, as done by Borexino. In that case, it is the fitting parameters of these functions, and not the survival probabilities themselves, that end up becoming the \lq\lq measured quantities\rq\rq\ in the inference framework, which adds another layer of complexity that we choose not to include here.}
}}.  From Borexino we used only the observed \textit{pp}-chain neutrinos (specifically, $^8$B neutrinos), and not the carbon-nitrogen-oxygen cycle (CNO) neutrinos. This is a reasonable choice for the Sun: its core temperature is relatively low, so that few CNO neutrinos are produced.  In addition, for simplicity we used daytime data only.  The Borexino survival probabilities are listed in Table~\ref{table:borex}.

\setlength{\tabcolsep}{5pt}
\begin{table}[htb]
\caption{Energy bins and corresponding survival probabilities $P_{ee,i}$ from the Borexino experiment~\cite{BOREXINO:2018ohr}.  These energy values were also adopted for the model.}
\vskip 0.2cm
\centering
\begin{tabular}{| c | c | c | c |} \toprule
\hline
 \textit{Parameter} & \textit{Energy [MeV]} & \textit{Parameter} & \textit{Probability} \\\midrule \hline
 $E_{1}$ & 7.4 & $P_{ee,1}$ & $0.39\pm0.09$\\
 $E_{2}$ & 8.1 & $P_{ee,2}$ & $0.37\pm0.08$\\
 $E_{3}$ & 9.7 & $P_{ee,3}$ & $0.35\pm0.09$ \\
 \bottomrule \hline
\end{tabular} 
\label{table:borex}
\end{table}

\section{THE INFERENCE METHOD} \label{sec:method}

This section offers a brief description of our methodology.  For details, we refer the reader to Ref.~\cite{laber2023inference}.  

\subsection{General formulation of SDA} \label{sec:methodGen}

SDA assumes that any observed quantities arise from an underlying dynamical model, and that those quantities represent only a sparse subset of the model's full degrees of freedom. We call this model $F_a(\bm{x}(r),\bm{p}(r))$; $a = 1,2,\ldots,D$: a set of $D$ ordinary differential equations governing the evolution of $D$ state variables $x_a(r)$, where $r$ is our parameterization and $\bm{p}$ are unknown parameters ($p$ in number).  

A subset $L$ of the $D$ state variables can be associated with measured quantities.  We seek to estimate the evolution of all $D$ state variables that is consistent with those $L$ measurements, and to predict their evolution at locations where measurements have not been obtained.

\subsection{A path integral approach} \label{sec:methodPathInt}

We can cast SDA as a path integral formulation, in the following sense.  We seek the probability of obtaining a path $\bm{X}$ given observations $\bm{Y}$: 
\begin{align*}
  P(\bm{X}|\bm{Y}) = e^{-A_0(\bm{X},\bm{Y})},
\end{align*}
\noindent 
which becomes a problem of minimizing the quantity $A_0$, our "action."  Further, we use an optimization formulation, where the cost function of the optimizer is equivalent to the action on a path in the state space.  The cost function surface is $((D + p) \times (N+1))$-dimensional, where $N+1$ is the number of discrete model locations, which we take to be independent dimensions.  We seek the path $\bm{X}^0 = \{ \bm{x}(0),\ldots,\bm{x}(N),\bm{p}(0),\ldots,\bm{p}(N) \}$ in state space that corresponds to the lowest cost.  We find minima via the variational method~\cite{oden2012variational}.

After many simplifications, $A_0$ can be written in the following computationally implementable form:
\begin{widetext}
\begin{equation} \label{eq:actionlong}
\begin{split}
A_0 =& R_f A_\text{model} + R_m A_\text{meas}\\
A_\text{model}=&\frac{1}{{N}D}	\mathlarger{\sum}_{n \in \{\text{odd}\}}^{N-2} \, \mathlarger{\sum}_{a=1}^D \\ 
   & \Bigg[ \left\{x_a(r_{n+2}) - x_a(r_n) - \frac{\delta r}{6} [F_a(\bm{x}(r_n), \bm{p}(r_n)) + 4F_a(\bm{x}(r_{n+1}),\bm{p}(r_{n+1})) + F_a(\bm{x}(r_{n+2}),\bm{p}(r_{n+2}))]\right\}^2 \\
   & + \left\{ x_a(r_{n+1}) - \frac12 \left(x_a(r_n)+x_a(r_{n+2})\right) - \frac{\delta r}{8} [F_a(\bm{x}(r_n),\bm{p}(r_n)) - F_a(\bm{x}(r_{n+2}),\bm{p}(r_{n+2)})]\right\}^2 \Bigg] \\
  A_{\text{meas}} =& \frac{1}{N_{\text{meas}}} \mathlarger{\sum}_{r_m \in \{\text{meas}\}} \, \mathlarger{\sum}_{l=1}^L  \left[\left(y_{l}\left(r_m\right) - h_{l,m}(\bm{x}(r_m) \right)^2 \right]
\end{split}
\end{equation}
\end{widetext}

In $A_\text{model}$, adherence to the model evolution is required of all $D$ state variables $x_a$.  The outer sum on $n$ runs through all odd numbered discretized locations.  The inner sum on $a$ runs through all $D$ state variables  { (for our case, these are polarization vector components [$P_{x,i},P_{y,i},P_{z,i}$] for each energy $E_i$)}.  The terms within the first and second sets of curly brackets represent the errors in the first and second derivatives, respectively, of the state variables.

In $A_\text{meas}$, we require adherence to the measurements of any measured quantities.  The variables $y_l$, for $l=1,\ldots,L$, are the $L$ components that are measured at locations $r_m \in \{{\text{meas}}\}$  { (for our case, these are the linear combinations of the components $P_{z,i}$  and $P_{x,i}$ of the polarization vectors for each energy $E_i$, in accordance with Eq.~\eqref{eq:connection}, at the final 1000 grid locations.)}   the number of locations is $N_\text{meas}$.  We will compare these values to the components $h_{l,m}(\bm{x})$.

These $h_{l,m}$ are transfer functions translating the model state variables to the measured quantities.   Here, the measured quantities are the values of $P_z$ of each neutrino, at two locations: the center of the Sun, and the surface of Earth (the "measurement" at the center of the Sun is really a robust theoretical expectation on neutrino flavor).  At the Sun's center, we can compare the measurement directly to the model's $P_z$, rendering the transfer functions trivial (that is: $h_0(\vec P) = P_z$.)  The translation at the surface of Earth, however, is more involved.  This is because our model grid does not extend beyond the Sun (i.e., beyond $r_\text{final} = R_\odot/2$), and the neutrinos experience kinematic decoherence on their way to Earth (see Sec.~\ref{sec:evol} and Ref.~\cite{laber2023inference} for an explanation).  We connect model to measurement at this end by comparing the $P_z$ measurement at Earth to an extrapolated $P_z$ value derived from the polarization vector components at $R_\odot/2$ [as given by Eq.~\eqref{eq:connection}].  In this way, there is an equivalency between measuring $P_z$ at Earth and measuring a linear combination of $P_z$ and $P_x$ at $R_\odot/2$. Thus, the transfer function at this end is:
\begin{equation} \label{eq:h}
    h_\text{final}(\vec P) = \left(1-\sin^22\theta\right) P_z- \cos2\theta \sin 2\theta \, P_x,
\end{equation}
for each neutrino energy. Then the measurement term becomes:

\begin{widetext}
\begin{equation} \label{eq:actionmeas}
\begin{split}
  A_{\text{meas}} =& \frac{1}{N_{\text{meas}}} \sum_{k=1}^{N_\nu} \left[\left(P_{z,k}^\text{meas} \left(0\right) - P_{z,k}\left(0\right)\right)^2 + \left(P_{z,k}^\text{meas}(\oplus) - h_\text{final}\left( \vec{P}_k\left(\frac{R_\odot}{2}\right) \right)\right)^2\right].
\end{split}
\end{equation}
\end{widetext}

Here, the subscript $k=\{1,\ldots,N_\nu\}$ indexes the neutrino energy bins. 
$P_{z,k}^\text{meas}$ is the measurement of $P_z$ at the specified location (with $0$ being the center of the Sun, and $\oplus$ being the Earth). { Table~\ref{table:connect} distills the relation between Eq.~\eqref{eq:actionlong} and the quantities of interest in the specific model of this paper.  For a detailed derivation of the cost function of Eq.~\eqref{eq:actionlong}, see Appendix~\ref{app:SDA}.}

\renewcommand{\arraystretch}{1.2}
\setlength{\tabcolsep}{5pt}
\begin{table*}
\caption{{Relation between language of Eq.~\eqref{eq:actionlong} and our model quantities.}}
\centering
\begin{tabular}{| c | c | c |} \toprule
\hline
 \textit{Eq.~\ref{eq:actionlong}} & \textit{Our model for inference} & \textit{Description} \\\midrule \hline
 $A_{model}$ & See Eq.~\eqref{eq:actionlong} & imposes dynamics of all state variables, measured and unmeasured \\
 $x_{a}$ & $P_{x,i}$, $P_{y,i}$, $P_{z,i}$ & all state variables \\
 $p$ & $V_0$ & unknown parameters \\\hline
 $A_{meas}$ & See Eq.~\eqref{eq:actionlong} & imposes measurements onto state variables associated with measured quantities \\
 $y_l$ & $P_{n,\text{survival}}$ & measured quantities\\
 $h_{l,m}$ & See Eq.~\eqref{eq:h}& translates between measured quantities and model state variables 
 \\\bottomrule \hline
\end{tabular} 
\label{table:connect}
\end{table*}
\renewcommand{\arraystretch}{1}


\subsection{\textbf{Multiple solutions}} \label{sec:methodAnnealing}

The action surface for a nonlinear model will be nonconvex, and our search algorithm is descent only.  Thus, we face the problem of multiple minima.  To identify a lowest minimum, we iteratively anneal~\cite{ye2015systematic} in terms of the coefficients $R_f$ and $R_m$ of the model and measurement error terms, respectively, of Eq.~\eqref{eq:actionlong}.  

We set $R_m$ to 1.0, and write $R_f$ as $R_{f,0}\alpha^{\beta}$.  The values of $R_{f,0}$ and $\alpha$ are chosen to work best for the particular model at hand (in this paper, they are $10^{3}$ and $2.0$, respectively), and $\beta$ is the annealing parameter, initialized at 1.  The first annealing iteration takes the measurement error to dominate: with the model dynamics imposed relatively weakly, the action surface is rather smooth, and we obtain an estimate of $A_0$.  The second iteration involves an integer increment in $\beta$, which places faint structure upon the action surface, and the search begins anew from the initial estimate of $A_0$.  We anneal toward the deterministic limit where $R_f \gg R_m$, aiming to remain sufficiently close to the lowest minimum along the way.

 

\section{SPECIFIC INFERENCE TASK} \label{sec:expers}



Our model contained neutrinos with three distinct energies, chosen to replicate the Borexino experiment's three energy bins.  For each neutrino energy, the flavor evolution was constrained by two measurements.  At a radius of {$r = 0$}, a measurement of \(P_z = 1\) was given to represent pure electron flavor.  At the outer end, $r = {R_\odot}/{2}$, \(P_z\) and \(P_x\) were constrained based on energy-dependent measurements of the survival probability \(P_{ee}\) at Earth, as described in Section~\ref{sec:methodPathInt}.  There we provided as measurements the final 1,000 pairs of those values; i.e. the 1,000th pair corresponded to the final radial location at $r = R_\odot/2$.  This choice was in keeping with our prior work (Ref.~\cite{laber2023inference}), and is justified given that the neutrinos had completed their matter-driven flavor evolution well before encountering this region in which the 1,000 measurements were made.  Knowing that the neutrinos decohere by the time they arrive at Earth, the relation [Eq.~\eqref{eq:connection} or Eq.~\eqref{eq:h}] between the measured $P_{ee}$ (or $P_z$) at Earth, and the $\{P_z, P_x\}$ pairs in the Sun, can be applied to any number of points within the "vacuum regime" of the domain.

We ran two versions of the optimization, each defined by a distinct choice for \(V(r)\) 
in the flavor evolution model [Eq.~\eqref{eq:model1}].  In one version, the model took the exponential form for \(V(r)\) [Eq.~\eqref{eq:expdecaymodel}].  The other version took the more complicated logistic form of [Eq.~\eqref{eq:logmodel}].  For each case, the value of \(V_f\) -- that is, the value at $R_\odot/2$ -- was taken to be a fixed known value of \(4.470945\times10^{-4}\,\text{ km}^{-1}\) (consistent with the expectation from the standard solar model). In comparison, the value of $\omega\cos{2\theta}$ at the typical Borexino energies is $\sim 0.01\,\text{km}^{-1}$, reinforcing our assertion that the endpoint of our domain is comfortably within the vacuum oscillation regime.

\subsection{\textbf{Proof of concept: State prediction at fixed $V_0$}}

Prior to performing parameter estimation, we needed to establish that the SDA procedure could reliably identify the correct solution for the case wherein the correct solution is not known to us.  Thus, we first sought to determine how well the procedure would predict state variable evolution in a controlled scenario wherein we did know the true value of $V_0$.  In this case, the predicted evolution of the state variables can be compared to the output from forward integration. 

In previous work~\cite{armstrong2020inference}, we showed that the value of the action could be used as a litmus test for the correct solution; namely, the correct solution corresponded to the path of least action.  Here, we sought to verify that that was indeed also the case for this model.  That is: the best match to the forward integration should also be the solution corresponding to the path of least action.  If we found that to be the case, then we would be able to identify the correct solution for the scenario wherein we do not have an independent verification from forward integration.  In short: the best solution is simply the one corresponding to the path of least action.  It is vital to have such a metric prior to trusting the data to lead us blind in parameter estimation.

\begin{figure}[htb]
    \centering
    \includegraphics[width=9cm]{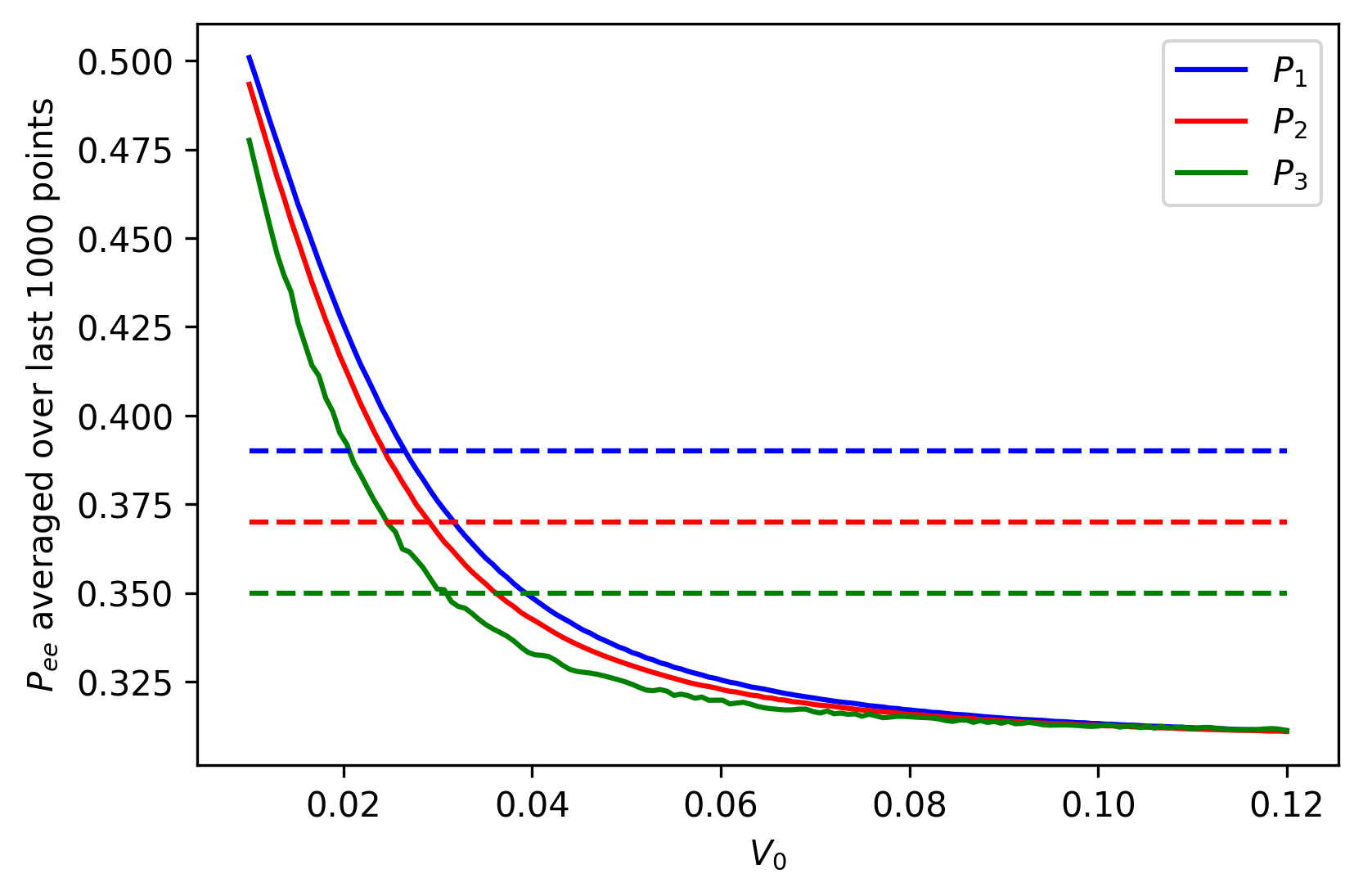}
    \caption{Predicted neutrino survival probabilities \(P_{ee}\), as would be measured on Earth based on Eq. \eqref{eq:connection}, versus $V_0$, the matter potential at the center of the Sun. These predictions were obtained using forward integration.  Dashed lines represent the measured survival probabilities for each neutrino energy in the Borexino experiment~\cite{BOREXINO:2018ohr}.  Here we used the logistic model for \(V(r)\) [Eq.~\eqref{eq:logmodel}]; results for the exponential decay model were visually identical.  We used this plot to determine our search range for $V_0$.  Importantly, this range spans the prediction from the standard solar model~\cite{bahcall2005new} of $V_0 \sim 0.03$\,km$^{-1}$.}
    \label{fig:logistic-V0P}
\end{figure}

To that end, for each model version, we performed eight variations, each taking a known distinct value of $V_0$.  We sought a range for these values that would encompass the prediction from the standard solar model~\cite{bahcall2005new} that $V_0$ is around 0.03 $\text{km}^{-1}$.  Further, we sought to identify the range over which the survival probabilities are expected to be sensitive to the value of $V_0$, using forward integration.  We examined \(V_0\) over the range \(\left[0.001,0.12\right] \text{km}^{-1}\), using 500 linearly spaced steps in $V_0$, each time initializing our state as \(\vec{P} = \left[0,0,1\right]\).  Then we calculated values of \(P_{ee}\) as they would be measured on Earth, via the transformation given in Eq.~\eqref{eq:connection}.  The outcome is shown in Fig.~\ref{fig:logistic-V0P}.  For both models of \(V(r)\), values of \(V_0\) near 0.025 \(\text{km}^{-1}\) to 0.030 \(\text{km}^{-1}\) produced \(P_{ee}\) values that most closely matched the measurements, for each neutrino energy. 
Given this outcome, we chose the following eight values for the proof-of-concept stage SDA experiments: \(0.015 \text{ km}^{-1},0.020 \text{ km}^{-1},0.025 \text{ km}^{-1},0.030 \text{ km}^{-1},\) \(0.035 \text{ km}^{-1},0.055 \text{ km}^{-1},0.075 \text{ km}^{-1}, \text{ and }0.095 \text{ km}^{-1}\). 

\subsection{\textbf{Parameter estimation of $V_0$}}

Once we ascertained that the path of least action reliably corresponded to the best state variable evolution match to the (known) forward integration result in that controlled scenario, and importantly that the reliability did not depend on the value of $V_0$ itself (see Sec.~\ref{sec:result}), we proceeded with the more ambitious problem of parameter estimation.  Specifically, at this stage we challenged the procedure to infer $V_0$, given the (now incomplete) model together with the Borexino data.  Here, the permitted search range for $V_0$ was: 0.001 to 0.120 km$^{-1}$, chosen to match the range used for the preliminary forward integration tests (Fig.~\ref{fig:logistic-V0P}).

We repeated the parameter estimation, this time considering the Borexino experimental error, by adding Gaussian-distributed noise\footnote{This noise was added to each of the 1,000 locations sampled for each energy.  \(0.1\%\) of the resulting noisy samples fell below the range \([0,1]\), which is unphysical.  So we imposed a lower bound at 0, which had negligible effect on the distribution.} of that average magnitude into the measurements of \(P_{ee}\) at Earth.  


We used the open-source interior-point optimizer (Ipopt)~\cite{wachter2009short} to perform the simulations.  Ipopt discretizes the state space via a Hermite-Simpson method, with constant step size.  We used 121,901 steps and a step size of $\delta r$ of 2.8524\,{km$^{-1}$}.  We discretized the state space, and calculated the model Jacobian and Hessian, using a Python interface~\cite{minAone} that generates C code that Ipopt reads.  The computing cluster that ran the simulations had 201 GB of RAM and 24 GenuineIntel CPUs (64 bits), each with 12 cores.

For the proof-of-concept test of 
the accuracy of the $V_0$ estimates, we generated a simulation via forward integration, with all neutrinos initialized at \(\left[P_x,P_y,P_z\right]=\left[0,0,1\right]\) at the solar core.  Here we used the value of $V_0$ obtained through parameter estimation, and a domain identical to the domain used in the optimization.
This result we compared directly to the solution from Ipopt.  The integration was performed by Python's odeINT package, which uses a FORTRAN library and an adaptive step size.  Our complete procedure can be found in the publicly available repository of Ref.~\cite{github}.

For all experiments, ten independent paths were initialized randomly.  That is, each initialization consisted of as many random choices as there are dimensions to the action surface: $D \times (N+1) + p$, where $D$, $N+1$, and $p$ are the number of state variables, discretized model locations, and parameters, respectively.  The permitted search range for state variables ($P_x$, $P_y$, and $P_z$) spanned their full dynamical range \(\left[-1.0,1.0\right]\).

\section{RESULTS AND DISCUSSION} \label{sec:result}


The key results are threefold: 
\begin{itemize}
    \item For the proof-of-concept experiments wherein $V_0$ was taken to be known; the Borexino data were most consistent with a model $V_0$ value in the range of 0.025-0.030 km$^{-1}$ (in keeping with the prediction from the standard solar model).  In addition, the state variable predictions that most closely matched the known result from forward integration were indeed the solutions corresponding to the paths of least action.  Thus we had confidence in using the path of least action as a litmus test in the parameter-estimation stage; a stage at which there would exist no independent "known truth" from forward integration.
    \item For the parameter estimations of $V_0$ based on the data: the path-of-least-action litmus test identified a (slightly broader) range for $V_0$ that centered around the (smaller) range identified above (of 0.025-0.030 km$^{-1}$), as most consistent with the Borexino measurements.  
    \item All results were invariant across the exponential and logistic models for \(V(r)\).  That is, it was not the shape of the matter profile that dictated the survival probability, but rather only the difference between initial ($V_0$) and final ($V_f$) values -- as expected for adiabatic flavor evolution.
\end{itemize}

\begin{figure*}[htb]
    \centering
    \includegraphics[width=17cm]{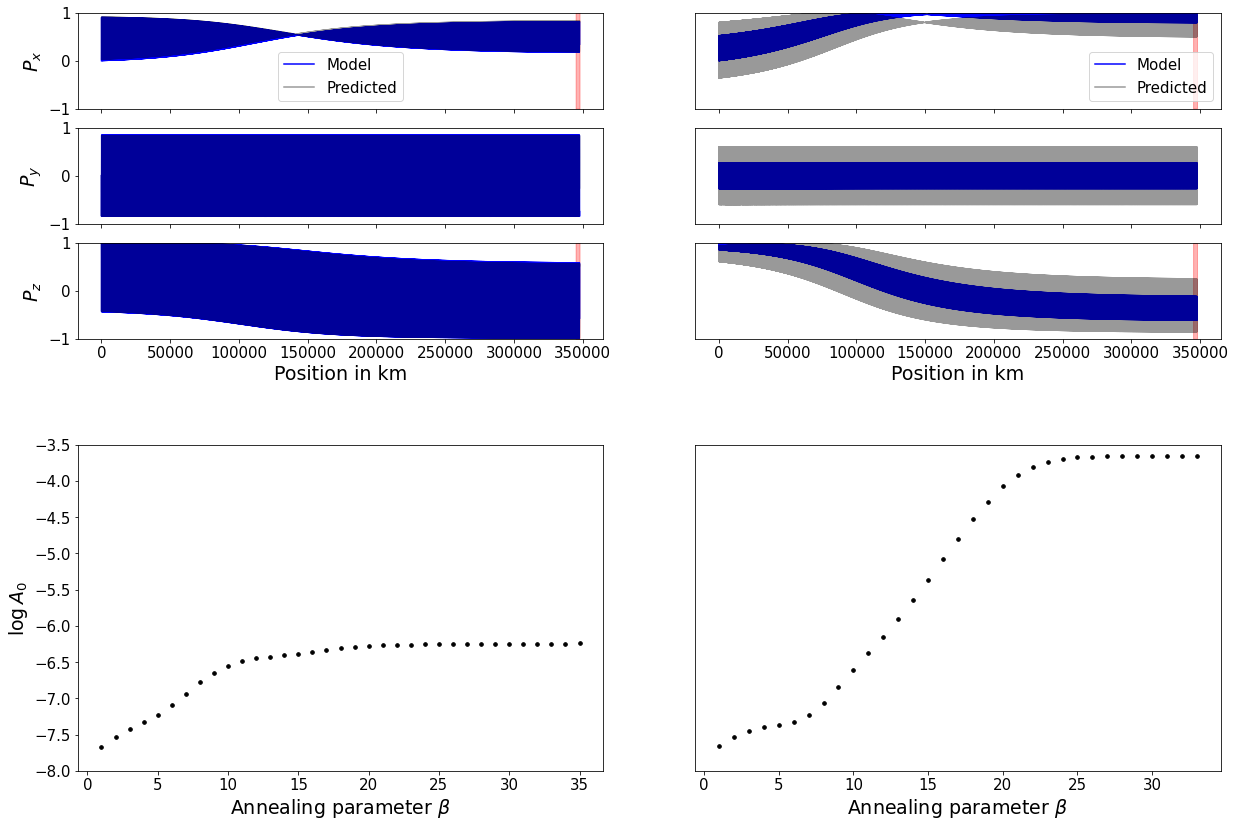}
    \caption{\textit{Top}: Evolution of state variables, using different fixed values of $V_0$ in each case: prediction (gray) versus expectation from forward integration (blue), for one neutrino energy, using the logistic model for $V(r)$.  \textit{Left}: \(V_0 = 0.025\) $km^{-1}$; \textit{right}: \(V_0 = 0.095\) $km^{-1}$.  The left match is significantly better, black versus blue are indiscernible by-eye.  \textit{Bottom}: Corresponding plots of log(action) versus annealing parameter $\beta$, for each result, respectively.  The action asymptotes to $\sim 10^{-6}$ (left) and $\sim 10^{-4}$ (right). Note that the better prediction (left) corresponds to the path of lower action.  The exponential model for $V(r)$ yielded similar results; not shown.  Thus the action can be taken as a metric for the case of parameter estimation, wherein we will have no independent check on prediction quality from forward integration (since $V_0$ will not be considered "known").  In addition, the transparent red band in the panels for $P_x$ and $P_z$ denote the (sparse) region where measurements of \(P_{ee}\) at Earth were provided, as per Eq.~\eqref{eq:connection}}. 
    \label{fig:statevar-0.025_noP}
\end{figure*}

\subsection{\textbf{Proof of concept: {State prediction at fixed $V_0$}}}
\label{sec:resultNoParam}

As described in Sec.~\ref{sec:expers}, our initial step -- prior to parameter estimation -- was to identify a metric that could be used to identify the optimal solutions for the case wherein the true evolution of state variables is not known.  To this end, we ran the optimization for the eight distinct (known) values of $V_0$, and compared the resulting predictions for state variables against the (known) results from forward integration.

The top panel of Fig.~\ref{fig:statevar-0.025_noP} shows the predicted evolution of state variables (black) compared to the result obtained from integration (blue), for one neutrino energy.  Left and right panels are examples of an excellent versus a poorer match, respectively.  In both, the overall flavor evolution is {qualitatively} predicted well, but the prediction on the right fails to capture the amplitude of oscillations. The difference between the procedures on the left versus right was as follows: on the left, the specific model value chosen for $V_0$ is 0.025\,km$^{-1}$; on the right, the choice is instead 0.095\,km$^{-1}$.  (This result was obtained using the logistic form for $V_0$; results are similar for the exponential form but are not shown.)  In other words, these results indicate that a value for $V_0$ of 0.025\,km$^{-1}$ is more compatible with the Borexino measurements, compared to the value of 0.095\,km$^{-1}$.  

Now, the bottom panel of Fig.~\ref{fig:statevar-0.025_noP} shows the corresponding plot of the action versus annealing parameter $\beta$, for each result, respectively.  The action for the better (left) solution asymptotes to a significantly lower value compared to that at right; namely, $A_0 \sim 10^{-6}$ compared to $10^{-4}$.  Indeed, this was our finding in a much more extensive study of the utility of this "path of least action" for identifying best solutions~\cite{armstrong2020inference}.  In addition, note the vertical red band on the plot for $P_x$ and $P_z$, denoting the region in which measurements were provided to the procedure -- as a reminder of their sparsity.

\begin{figure}[htb]
    \includegraphics[width=9cm]{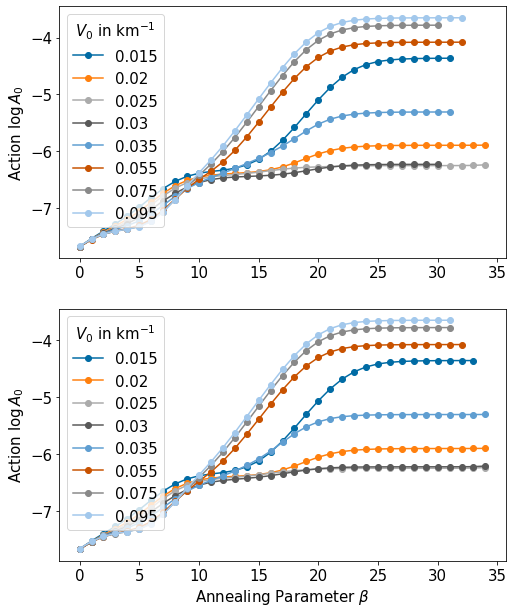}
    \caption{Log of action as a function of annealing parameter $\beta$ for each fixed value of \(V_0\) chosen {in the proof-of-concept stage of optimization}, for the logistic [Eq.~\eqref{eq:logmodel}] model for \(V(r)\) (\textit{top}), and the exponential [Eq.~\eqref{eq:expdecaymodel}] model (\textit{bottom}).  For both models, values of 0.025 and 0.03\,km$^{-1}$ resulted in the lowest action, with either lower or higher values of $V_0$ corresponding to a higher action plateau. }
    \label{fig:log-and-exp-action_noP}
\end{figure}

We repeated the comparison shown in Fig.~\ref{fig:statevar-0.025_noP} across all chosen values for $V_0$.  Resulting composite action plots are shown in Fig.~\ref{fig:log-and-exp-action_noP}, for the logistic (top panel) and exponential (bottom) versions.  Note that the greater the deviation of the chosen value of $V_0$ from the expected range of [0.025, 0.030]\,km$^{-1}$, the higher the action value.  And critically: the higher the action value, the poorer the optimization of state variable evolution with the data (not shown).  Thus, within the scope of this paper, the action can be taken as a reliable proxy for the correct solution, for the case wherein we "fly blind" without an independent check from forward integration.

\subsection{\textbf{Parameter estimation of $V_0$}} \label{sec:resultParam}

With our action-as-litmus-test in hand, we proceeded with parameter estimation of $V_0$, given the Borexino data.  As per the description in Sec.~\ref{sec:expers}, we performed ten independent trials for each model version of $V(r)$, once without noise, and once with Gaussian noise on the order of the published Borexino errors~\cite{BOREXINO:2018ohr} added to the measurements.

\begin{figure}[htb]
    \centering
    \includegraphics[width=9cm]{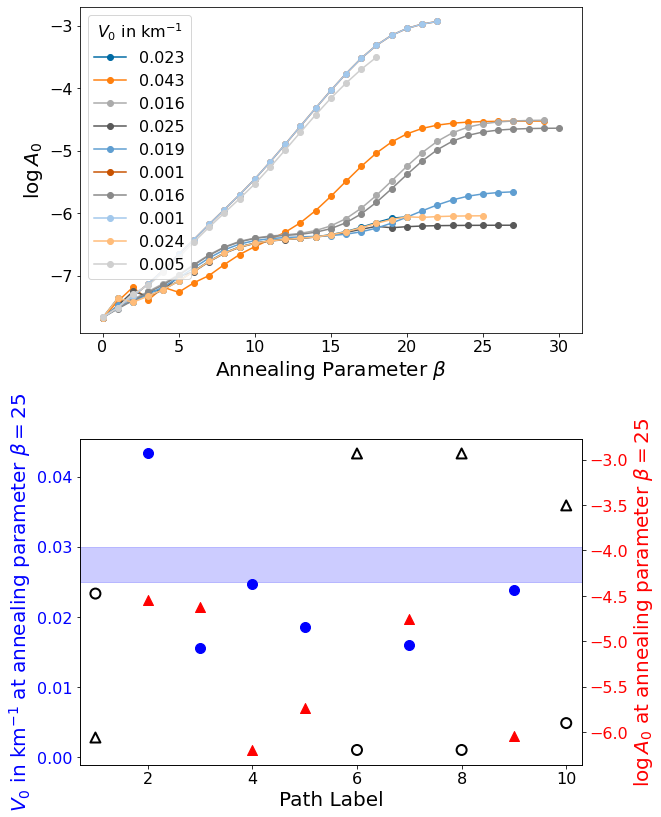}
    \caption{\textit{Top}: Log of action as a function of annealing parameter $\beta$ for the logistic model, across ten independent trials, without noise in measurements. The values of $V_0$ in the plot legend correspond to a value of 25 for annealing parameter $\beta$; an iteration at which solutions had stabilized.  \textit{Bottom}: Estimate of \(V_0\) [left (blue) $y$-axis, circles] and \(\log A_0\) at \(\beta = 25\) [right (red) $y$-axis, triangles], across the ten trials.  The light blue band indicates the expected range of \(V_0\), based on Fig.~\ref{fig:logistic-V0P} and on the proof-of-concept optimization runs.  (Those trials that did not plateau are denoted by hollow shapes, both for the $V_0$ estimates and for the corresponding value of the action.)}
    \label{fig:logistic-param-results}
\end{figure}

\begin{figure}[htb]
    \centering
    \includegraphics[width=9cm]{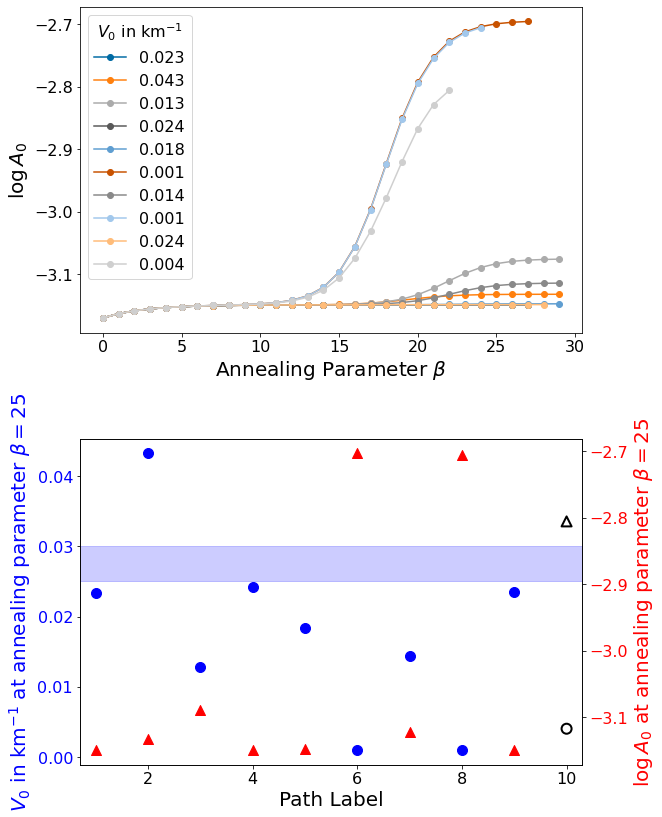}
    \caption{Same as Fig.~\ref{fig:logistic-param-results} for the logistic model, but showing the results for the optimization trials with Gaussian noise added to the measurements, on the order of the uncertainties in the Borexino data~\cite{BOREXINO:2018ohr}.
    }
    \label{fig:logistic-param-noisy}
\end{figure}

Results for the logistic model are shown in Fig.~\ref{fig:logistic-param-results} (noiseless) and Fig.~\ref{fig:logistic-param-noisy} (noisy).  The top panel of each plot shows the log of the action as a function of $\beta$, for all ten trials.  All trials found stable estimates.  With the noiseless data, the estimates of \(V_0\) spanned the range \([0.0105, 0.0488]\text{ km}^{-1}\), where the lowest value of the action \(A_0\) corresponded to an estimate for \(V_0\) of 0.0247 \(\text{ km}^{-1}\).  With additive noise, the range for \(V_0\) was \([0.001, 0.0432]\text{ km}^{-1}\), and the value corresponding to lowest action was \(V_0 = 0.0243 \text{ km}^{-1}\). 

The bottom panels of Figs.~\ref{fig:logistic-param-results} and \ref{fig:logistic-param-noisy} offer an alternative display of the information contained in the top panel.  Here, the estimate of $V_0$ is shown on the left (blue) $y$-axis, and the corresponding asymptotic value of the action is on the right (red) $y$-axis.  The horizontal light blue band indicates the expected range for $V_0$, based on the analysis depicted in Fig.~\ref{fig:logistic-V0P} or Fig.~\ref{fig:log-and-exp-action_noP}.  Note that the greater the distance of the estimate of $V_0$ from that band (blue circle), the higher the action value (red triangle).  (Paths whose action value did not reach a stable plateau are denoted with hollow shapes.)

\begin{figure}[htb]
    \centering
    \includegraphics[width=9cm]{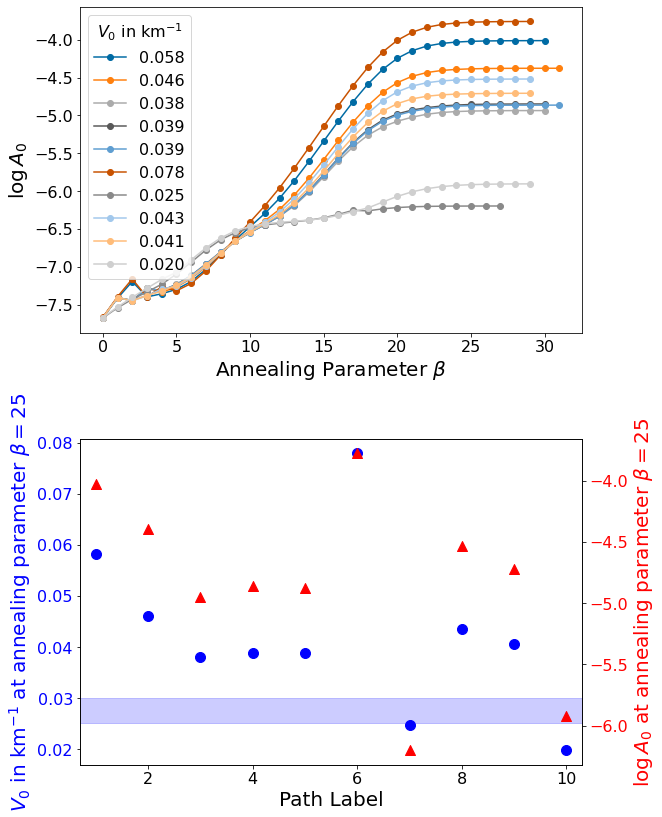}
    \caption{Same as Fig.~\ref{fig:logistic-param-results}, but with the exponential-decay model for $V(r)$ [Eq.~\eqref{eq:expdecaymodel}] instead of the logistic model [Eq.~\eqref{eq:logmodel}], for noiseless measurements.
    }
    \label{fig:expo-param-results}
\end{figure}

\begin{figure}[htb]
    \centering
    \includegraphics[width=9cm]{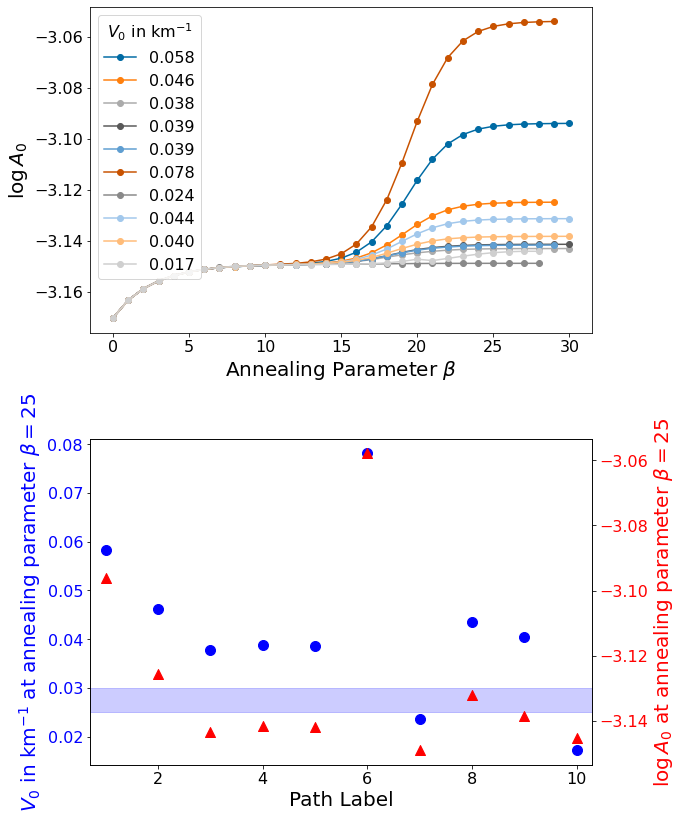}
    \caption{Same as Fig.~\ref{fig:expo-param-results} for the exponential-decay model for $V(r)$ [Eq.~\eqref{eq:expdecaymodel}], with Gaussian noise added to the measurements, on the order of the Borexino uncertainties~\cite{BOREXINO:2018ohr}
    }
    \label{fig:expo-noise-param-results}
\end{figure}

Results for the exponential model were similar.  Figures~\ref{fig:expo-param-results} and ~\ref{fig:expo-noise-param-results} show the noiseless and noisy cases, respectively.  The without-noise range of estimates for \(V_0\) was \(\left[0.0198, 0.0779\right]\,\text{km}^{-1}\), and with noise the range shifted slightly to \(\left[0.0173, 0.0781\right]\,\text{km}^{-1}\).  Again, for both noiseless and noisy cases, the path of least action corresponded to a value for \(V_0\) within the expected range of \(0.025\text{--}0.030\)\,km$^{-1}$.

\section{CONCLUDING REMARKS}

To summarize our results: neutrino flavor measurements at terrestrial detectors (e.g., Borexino) can be used to independently constrain the electron density at the solar center (complementary to the existing constraints from stellar oscillations, for instance). This is demonstrated here through a novel application of the SDA method. The outcome is insensitive to the overall shape of the electron density profile, as long as it evolves sufficiently smoothly through the mass level crossing between the instantaneous in-medium neutrino mass eigenstates -- ensuring adiabatic flavor evolution across the resonance.

{ One reason we are interested in independent means to estimate the electron density profile $n_e(r)$ is that $n_e(r)$ can inform other important solar properties.  For example, there exists a strong connection between helioseismology and neutrino production in the Sun~\cite{RevModPhys.74.1073} -- a connection that can be exploited. 
Specifically, taken together with helioseismology data, $n_e(r)$ can lead to an estimate of the solar temperature profile.  The connection is sound speed.  Observations of pressure waves at the solar surface have grown increasingly precise for deducing the sound speed profile throughout the Sun and sound speed at any given location in the Sun depends on both matter density and temperature at that location.}

{ As another future direction,} it is worthwhile to investigate possible improvements that inference might offer for the analysis of solar neutrino data { itself}.  For example, it might provide an independent check on new methods to remove cosmogenic-induced spallation in Super-Kamiokande, a recent effort to improve the precision of solar neutrino data~\cite{Super-Kamiokande:2021snn}.  

{ Finally, and looking beyond our immediate solar neighborhood; the inverse paradigm offers a unique and possibly complementary counterpoint to integration, for studying a diversity of astrophysical environments from "simple" main sequence evolution to the fiercely nonlinear environs of core-collapse supernovae and merger events.}

\section{ACKNOWLEDGMENTS}

C.~L., E.~A., L.~N., S.~R., and H.~T. acknowledge an Institutional Support for Research and Creativity grant from New York Institute of Technology, and NSF Grants PHY-2139004 and PHY-2310066.  E.~K.~J. acknowledges the NSF summer Research Experience for Undergraduates program under NSF Grant EAR-2050852.  The research of A.~B.~B. was supported in part by the U.S. National Science Foundation Grants No. PHY-2020275 and PHY-2108339.  A.~V.~P. acknowledges support from the U.S. Department of Energy (DOE) under contract number DE-AC02-76SF00515 at SLAC National Accelerator Laboratory and DOE grant DE-FG02-87ER40328 at the University of Minnesota. E.~A., A.~B.~B., and A.~V.~P. thank the Mainz Institute for Theoretical Physics (MITP) of the Cluster of Excellence PRISMA+ (Project ID No. 39083149) for its hospitality and support.  All coauthors extend gratitude to the good people of Kansas and Doylestown, Ohio.

\appendix
{

\section{DERIVATION OF THE COST FUNCTION}\label{app:SDA}

\subsection{Summary}

To derive the cost function, or "action", of Eq.~\eqref{eq:actionlong}, we first seek the probability of obtaining a path $\bm{X}$ in a model's state space given observations $\bm{Y}$: $P(\bm{X}|\bm{Y})$,
\begin{equation*}
  P(\bm{X}|\bm{Y}) = e^{-A_0(\bm{X},\bm{Y})}.
\end{equation*}
\noindent This is the path $\bm{X}$ for which the probability -- given $\bm{Y}$ -- is highest is the path that minimizes $A_0$.  If $A_0$ is sufficiently large (where "large" is defined within the context of a particular model), we can use Laplace's method to estimate the minimizing path on $A_0$\footnote{Laplace’s method was developed to approximate integrals of the form: $\int e^{Mf(x)}dx$.  For large $M$, significant contributions to the integral come only from points near the minimum.}.    

Then the expectation value of any function $G(\bm{X})$ on a path $\bm{X}$ can be written as:
\begin{align*}
  \langle G(\bm{X}) \rangle &= \frac{\int d\bm{X} G(\bm{X}) e^{-A_0(\bm{X},\bm{Y})}}{\int d\bm{X} e^{-A_0(\bm{X},\bm{Y})}}.
\end{align*}   
\noindent That is, the expectation value is written as a weighted sum over all paths, with weights exponentially sensitive to $A_0$\footnote{The rms variation and higher moments can be calculated by taking the $x_a$ to the appropriate higher exponents.}.  In our case, the quantity of interest is the path $\bm{X}$ itself.  

We now seek a functional form for $A_0$.  As a reminder of the notation the model consists of $D$ partial differential equations (PDEs), written in $D$ state variables.  We are able to measure $L$ quantities, each of which corresponds to one of the model's $D$ state variables.  Typically the measurements are sparse ($L \ll  D$), and the sampling may be sparse as well. 

\subsection{Model dynamics alone, without measurements}

We first consider the model's dynamics in the absence of measurements.  We represent the model's path through state space as the set $\bm{X} = \{\bm{x}(r_0),\bm{x}(r_1),\ldots,\bm{x}(r_N),\bm{p}\}$, where we have chosen a parametrization of $r$ (or radius).  Here, $r_N$ is the final discretized location and the vector $\bm{x}(r)$ contains the values of the $D$ total state variables, and $\bm{p}$ are the unknown parameters.  

\subsubsection{Assuming that a Markov process underlies the dynamics}

If we assume that the dynamics are Markov, then $\bm{x}(r)$ is determined by $\bm{x}(r-\Delta r)$, where $r-\Delta r$ means: "the location immediately preceding $r$" and an appropriate discretization of time $\Delta r$ for our particular model has been chosen.  A Markov process can be described in terms of differential equations,
\begin{align*}
  \diff{x_a(r)}{r} &= F_a(\bm{x}(r),\bm{p}); \hspace{1em} a =1,2,\ldots,D,
\end{align*}
with state variables $\bm{x}(r)$ and unknown parameters $\bm{p}$.  In this appendix we take the parameters to be constant numbers, but more generally they may vary with discretization $r$.

To discretize that continuous form, we choose to use the trapezoidal rule,
\begin{align*}
  x_a(n+1) = x_a(n) + \frac{\Delta r}{2}[F_a(\bm{x}(n+1)) +  F_a(\bm{x}(n))],
\end{align*}
where $n$ and $n+1$ are shorthand for $r_n$ and $r_{n+1}$.

\subsubsection{Introducing stochasticity}

We now permit stochasticity in the model dynamics.  Here the evolution can be written in terms of transition probabilties: $P(\bm{x}(n+1)|\bm{x}(n))$ is the probability of the system reaching a particular state at location $n+1$ given its state at location $n$.  If the process were deterministic, then in our case $P(\bm{x}(n+1)|\bm{x}(n))$ would reduce to: $\delta^D (\bm{x}(n+1) - \bm{x}(n) - \frac{\Delta r}{2}\left[\bm{F}(\bm{x}(n+1)) + \bm{F}(\bm{x}(n))\right])$.  We will revisit to this expression later.

For a Markov process, the transition probability from state $\bm{x}(n)$ to state $\bm{x}(n+1)$ represents the probability of reaching state $\bm{x}(n+1)$ given $\bm{x}(n)$ and $\bm{x}$ at all prior steps, or,
\begin{align*}
  P(\bm{x}(n+1)|\bm{x}(n)) &= P(\bm{x}(n+1)|\bm{x}(n),\bm{x}(n-1),\ldots,\bm{x}(0))
\end{align*}
so that
\begin{align*}
  P(\bm{X}) &\equiv P(\bm{x}(0), \bm{x}(1),\ldots, \bm{x}(N)) \\
			&= \mathlarger \prod_{n=0}^{N-1} P(\bm{x}(n+1)|\bm{x}(n)) P(\bm{x}(0)).
\end{align*}

Then we write 
\begin{align*} 
  P(\bm{X}) \equiv e^{-A_0(\bm{X})}, 
\end{align*}
so that the path $\bm{X}$ that minimizes the action $A_0$ is the path most likely to have occurred.  Then the model term of the action can be written as
\begin{align*}
  A_{0,\text{model}} = -\mathlarger{\sum} \log[P(\bm{x}(n+1)|\bm{x}(n))];
\end{align*}  
\noindent

\subsection{Adding measurements}

We now define measurements $\bm{Y}$ as the set of all vectors $\bm{y}(n)$ at all locations $n$, the analog of $\bm{X}$ for the state variables.  We shall examine the effect these measurements have on the model's dynamics by invoking "conditional mutual information" (CMI)~\cite{wyner1978definition}\footnote{For an intuition regarding CMI, consider that the overall information, in bits, in a set $A$ is defined as the Shannon entropy $H(A) = -\sum_A P(A) \log[P(A)]$.  The CMI quantifies the amount of information, in bits, that is transferred within a particular segment along a model trajectory.  That information is: $-\mathlarger \sum_{n=0}^N \log[P(\bm{x}(n)|\bm{y}(n),\bm{Y}(n-1))]$.}.  

The expression $\text{CMI}(\bm{x}(n),\bm{y}(n)|\bm{Y}(n-1))$ asks "how much is learned about event $\bm{x}(n)$  upon observing event $\bm{y}(n)$, conditioned on having observed event(s) $\bm{Y}(n-1)$?"  We can write it as
\begin{align*}
\text{CMI}(&\bm{x}(n),\bm{y}(n)|\bm{Y}(n-1)) \\
&= \log\left[\frac{P(\bm{x}(n),\bm{y}(n)|\bm{Y}(n-1))}{P(\bm{x}(n)|\bm{Y}(n-1)) P(\bm{y}(n)|\bm{Y}(n-1))}\right].
\end{align*}

\subsection{Action including measurements}

With measurements, the action becomes,
\begin{align*}
  A_0(\bm{X},\bm{Y}) = -\mathlarger{\sum} \log[P(\bm{x}(n+1)|\bm{x}(n))] - \log[P(\bm{x}(0))] \\
  - \mathlarger{\sum} \text{CMI}(\bm{x}(n),\bm{y}(n)|\bm{Y}(n-1)),
\end{align*}  
\noindent
where the first term represents the model dynamics, and the third term represents the transfer of information from measurements\footnote{The measurement term can be considered to be a synchronization term.  While such terms are often introduced artificially in optimization and control problems, we have shown that the measurement term can arise naturally through considering the effects of the information contained in those measurements.  In the absence of measurements, we live in a state space restricted only by our model degrees of freedom.  The measurements guide us to a solution within a \textit{sub}space in which those measurements are possible.
}.  The summations are over the discretized parameterization, which -- in this paper -- is a one-dimensional distance.  As noted, this formulation positions us to calculate the expectation value of any function $G(\bm{X})$ on the path $\bm{X}$.  

\subsection{Writing a calculable form for the action}

We now simplify the action formulation in order to obtain a form that can be implemented computationally.

\subsubsection{The measurement term}

Regarding the measurement term, we make four assumptions:
\begin{itemize}
  \item The measurements taken at different locations along the parameterized path are independent of each other.  Then the CMI term is $P(\bm{x}(n)|\bm{y}(n))$ or
  \begin{equation*}
    A_0(\bm{X},\bm{Y}) = -\log[P(\bm{X}|\bm{Y})].
  \end{equation*}
  \item There may be an additional relation between the measurements and the state variables to which those measurements correspond, which can be expressed via a transfer function $h_l$: $h_l(\bm{x}(n)) = y_l(n)$.
  \item For each of the $L$ measured state variables, we allow for noise $\theta_l$ at each measurement location, for each measurement $y_l$: $y_l(n) = h_l(\bm{x}(n)) + \theta_l(n)$.  Then $P(\bm{x}(n)|\bm{y}(n))$ is simply some function of $h(\bm{x}(n)) - \bm{y}(n)$ at each location.
  \item The measurement noise assumes a Gaussian distribution.
\end{itemize}  
\noindent
Then we arrive at:
\begin{align*}
  &\text{CMI}(\bm{x}(n),\bm{y}(n)|\bm{Y}(n-1)) \\
	&= -\mathlarger{\sum}_{l,k=1}^L (h_l(\bm{x}(n)) - y_l(n)) \frac{[R_m(n)]_{lk}}{2}(h_k(\bm{x}(n)) - y_k(n)),
\end{align*}
\noindent
where $R_m$ is the inverse covariance matrix of the measurements $y_l$.

\subsubsection{The model term}

We assume that the model may have errors, which will broaden the delta function in the expression noted earlier for the deterministic case.  We assume that the distribution of errors is Gaussian, so $\delta^D(\bm{z})$ becomes: $\sqrt{\mathlarger{\frac{\det R_f}{(2\pi)^D}}} \mathlarger{e^{\mathlarger{[-\bm{z} \frac{R_f}{2} \bm{z}]}}}$.  Here, $R_f$ is the inverse covariance matrix for the state variables.

Taking all approximations together, we can arrive at the full form of Eq.~\eqref{eq:actionlong}, with details in the main text of Sec.~\ref{sec:methodPathInt}.  For  complete details of this derivation, see Ref.~\cite{abarbanel2013predicting}.

}

{
\section{NEUTRINO PROPAGATION FROM SUN TO EARTH} \label{app:decoh}

\subsection{Coherent regime (solar surface)}
A neutrino produced in the center of the Sun goes through MSW resonance and eventually enters the vacuum oscillation regime. In the coherent regime (e.g., just outside the surface of the Sun), the density matrix in the mass basis in vacuum can be parameterized as
\begin{equation}
    \rho_{\nu,\odot}^{(m)} = 
    \begin{bmatrix}
        n_{\nu_1} & \rho_{12} \\
        \rho_{12}^* & n_{\nu_2}
    \end{bmatrix},
\end{equation}
where $n_{\nu_1}$ and $n_{\nu_1}$ are independent of time (distance) in the absence of matter effects, and $\rho_{12}$ has time-dependent oscillation phases. In the flavor basis, the same matrix can be parameterized as
\begin{equation}
    \rho_{\nu,\odot}^{(f)} = 
    \begin{bmatrix}
        n_{\nu_e} & \rho_{ex} \\
        \rho_{ex}^* & n_{\nu_x}
    \end{bmatrix},
\end{equation}
where all of the matrix components now depend on time, on account of the change of basis transformation described below. This transformation is given by the unitary matrix
\begin{equation}
    U = 
    \begin{bmatrix}
        \cos\theta & \sin\theta \\
        -\sin\theta & \cos\theta
    \end{bmatrix},
\end{equation}
so that $\ket{\nu_\alpha} = \sum_i U_{\alpha i}^* \ket{\nu_i}$. With this transformation, we can relate the density matrices in the two bases as $\rho^{(f)} = U \rho^{(m)}U^\dagger$. Therefore, one has
\begin{widetext}
\begin{equation}
    \begin{split}    
    \rho_{\nu,\odot}^{(f)} &= 
    \begin{bmatrix}
        n_{\nu_e} & \rho_{ex} \\
        \rho_{ex}^* & n_{\nu_x}
    \end{bmatrix}
    =
    \begin{bmatrix}
        \cos\theta & \sin\theta \\
        -\sin\theta & \cos\theta
    \end{bmatrix}
    \begin{bmatrix}
        n_{\nu_1} & \rho_{12} \\
        \rho_{12}^* & n_{\nu_2}
    \end{bmatrix}
    \begin{bmatrix}
        \cos\theta & -\sin\theta \\
        \sin\theta & \cos\theta
    \end{bmatrix} \\
    &= 
    \begin{bmatrix}
        n_{\nu_1} \cos^2\theta + n_{\nu_2}\sin^2\theta + \Re\rho_{12}\sin{2\theta} & \frac12(n_{\nu_2} - n_{\nu_1})\sin2\theta + \rho_{12}\cos^2\theta - \rho_{12}^*\sin^2\theta \\
        \frac12(n_{\nu_2} - n_{\nu_1})\sin2\theta + \rho_{12}^*\cos^2\theta - \rho_{12}\sin^2\theta & n_{\nu_1} \sin^2\theta + n_{\nu_2}\cos^2\theta - \Re\rho_{12}\sin{2\theta}
    \end{bmatrix}.
    \end{split}
    \label{eq:rhonufl}
\end{equation}
\end{widetext}
Note how the time dependence of $\rho_{12}$ appears in both the diagonal and off-diagonal components of the flavor-basis density matrix. In the polarization vector language, this becomes:
\begin{align}
    P_{z,\odot}^{(f)}  &=  n_{\nu_e} - n_{\nu_x} = (n_{\nu_1} - n_{\nu_2})\cos2\theta + 2\Re\rho_{12}\sin2\theta \nonumber \\
    P_{x,\odot}^{(f)}  &=  2\Re\rho_{ex} =  -(n_{\nu_1} - n_{\nu_2})\sin2\theta + 2\Re\rho_{12}\cos 2\theta \nonumber \\
    P_{y,\odot}^{(f)}  &=  -2\Im\rho_{ex} = -2\Im\rho_{12} \label{eq:polsun}
\end{align}


\subsection{Incoherent regime (earth)}

Next, we look at the incoherent regime, e.g., when the neutrinos arrive at the earth. By this point, the different mass eigenstates comprising a neutrino wave function have become spatially separated due to different velocities, and as a result, the neutrinos are detected at the earth as incoherent mixture of mass eigenstates. This phenomenon is called kinematic decoherence. In this state, the neutrino density matrix is given by:
\begin{equation}
    \rho_{\nu,\oplus}^{(m)} = 
    \begin{bmatrix}
        n_{\nu_1} & 0 \\
        0 & n_{\nu_2}
    \end{bmatrix},
\end{equation}
in the mass basis. The flavor basis density matrix $\rho_{\nu,\oplus}^{(f)}$ can then be obtained by setting $\rho_{12} \rightarrow 0$ in Eq.~\eqref{eq:rhonufl}. Now we see that there is no time dependence in $\rho_{\nu,\oplus}^{(f)}$, which also makes it easier to interpret the result of any measurement (since we no longer need to worry about any oscillating quantities). In this limit, the polarization vector components in the flavor basis are:
\begin{align}
    P_{z,\oplus}^{(f)} &=  (n_{\nu_1} - n_{\nu_2})\cos2\theta \nonumber \\
    P_{x,\oplus}^{(f)} &= -(n_{\nu_1} - n_{\nu_2})\sin2\theta \nonumber \\
    P_{y,\oplus}^{(f)} &= 0 \label{eq:polearth}
\end{align}

Comparing Eqs.~\eqref{eq:polsun} and \eqref{eq:polearth}, one can obtain the following relation:
\begin{equation}
    P_{z,\odot}^{(f)} \cos2\theta - P_{x,\odot}^{(f)} \sin2\theta = P_{z,\oplus}^{(f)} \cos2\theta - P_{x,\oplus}^{(f)} \sin2\theta.
\end{equation}

From Eqs.~\eqref{eq:polearth}, we also have
\begin{equation}
    P_{x,\oplus}^{(f)} = - P_{z,\oplus}^{(f)} \sin2\theta/\cos2\theta,
\end{equation}
and hence we can write: 
\begin{equation}
\begin{aligned}
    P_{z,\odot}^{(f)} \cos2\theta - P_{x,\odot}^{(f)} \sin2\theta = P_{z,\oplus}^{(f)}/\cos2\theta \\ 
    = (2P_{ee} - 1)/\cos2\theta,
    \label{eq:sunearthtrans}
\end{aligned}
\end{equation}
where $P_{ee}$ is the electron flavor survival probability measured at the detectors. Equation~\eqref{eq:sunearthtrans} represents the transformation between the neutrino state variables $P_{z,\odot}^{(f)}$ and $P_{x,\odot}^{(f)}$ in the outer parts of the Sun (where the matter density is low enough to essentially be in the vacuum oscillation regime) and the measured survival probability $P_{ee}$ at the earth.
}

%


\end{document}